# Attributes of a Great Requirements Engineer

Larissa Barbosa*, Sávio Freire†, Rita S. P. Maciel*, Manoel Mendonça*,
Marcos Kalinowski‡, Zadia Codabux§ and Rodrigo Spínola¶

*Federal University of Bahia, Salvador, Bahia, Brazil
Email: {larissa.leoncio, rita.suzana, manoel.mendonca}@ufba.br
†Federal Institute of Ceará, Morada Nova, Ceará, Brazil
Email: savio.freire@ifce.edu.br
‡Pontifical Catholic University of Rio de Janeiro, Rio de Janeiro, Brazil
Email: kalinowski@inf.puc-rio.br
§University of Saskatchewan, Saskatoon, Canada
Email: zadiacodabux@ieee.org
¶Virginia Commonwealth University, Richmond, VA, United States
Email: spinolaro@vcu.edu

*Abstract*—[Context and Motivation] Several studies have investigated attributes of great software practitioners. However, the investigation of such attributes is still missing in Requirements Engineering (RE). The current knowledge on attributes of great software practitioners might not be easily translated to the context of RE because its activities are, usually, less technical and more human-centered than other software engineering activities. [Question/Problem] This work aims to investigate which are the attributes of great requirements engineers, the relationship between them, and strategies that can be employed to obtain these attributes. We follow a method composed of a survey with 18 practitioners and follow up interviews with 11 of them. [Principal Ideas/Results] *Investigative ability in talking to stakeholders*, *judicious*, and *understand the business* are the most commonly mentioned attributes amongst the set of 22 attributes identified, which were grouped into four categories. We also found 38 strategies to improve RE skills. Examples are *training*, *talking to all stakeholders*, and *acquiring domain knowledge*. [Contribution] The attributes, their categories, and relationships are organized into a map. The relations between attributes and strategies are represented in a Sankey diagram. Software practitioners can use our findings to improve their understanding about the role and responsibilities of requirements engineers.

## I. INTRODUCTION

Several studies have investigated attributes of great software practitioners [1]–[7]. For instance, Li et al. [4] identified 53 attributes of great software engineers, such as adaptable, knowledge about people and the organization, and creativity. The authors also divided the attributes into categories, revealing that the attributes can be either internal, i.e., related to engineer's personality and ability to make effective decisions, or external, i.e., associated with the impact that great engineers have on people and software products. Another work investigated attributes of great software maintainers of open-source projects, revealing that technical excellence and communication are the most recurring attributes [6]. Knowing these results is necessary to help growing better software engineers and maintainers in this ever-demanding software industry.

Nevertheless, we believe current information about the attributes of great software practitioners might not be easily translated to the context of requirements engineering (RE). RE is a human intensive area. While several software engineering activities such as analysis, design, coding, and testing tend to make use of automated processes, activities related to RE are more related to social interactions among stakeholders involved in software development processes [8]. Requirements engineers are "a translator that understands the domain as well as its particular language well enough and also possesses enough IT know-how to be aware of the problems the developers face and to be able to communicate with them on the same level" [8].

In the book *Requirements Engineering Fundamentals*, Pohl and Rupp [8] mentioned that analytic thinking, empathy, communication skills, conflict resolution skills, moderation skills, self-confidence, and persuasiveness are expected attributes of a requirements engineer. Another text book of the area, *Handbook for the Advanced Level Requirements Elicitation* [9], describes the following attributes: communication, courtesy, flexibility, integrity, interpersonal skills, positive attitude, professionalism, responsibility, teamwork, and work ethic. These attributes, somehow, support our guess that attributes of requirements engineers have their own particularities. However, in both text books, there is a lack of empirical evidence supporting the described list of attributes. Indeed, to the best of our knowledge, there is not an investigation on attributes of a great requirements engineer. The lack of empirical evidence, in our understanding, hinders researchers from reasoning about them, employers from identifying them, and young engineers from becoming them.

Software RE is a determinant of productivity and product quality and it is inherently complex due to influences from customer environment, making the requirements engineer a central role in software development processes [10], [11]. There is a need to understand which attributes of requirements engineers are considered relevant in software development organizations in present day and how to pursue them. This is especially relevant, for instance, from the perspective of graduates and other novice software engineers seeking and recruited to requirements engineering jobs. Better understanding of the current desirable attributes can also be utilized in revising and improving RE education.

In this paper, we investigate the attributes of great requirements engineers and the relationship between them. To guide our research, we designed the following research questions (RQs):

- **RQ1**: What are the attributes of a great requirements engineer?
- **RQ2**: How these attributes are related to each other?
- **RQ3**: What strategies can be used to pursue these attributes?
- **RQ4:** How practitioners perceive the importance of the identified strategies to pursue the attributes of great requirements engineers?

We use a research method composed of a survey and follow up interviews. Through the survey, we ask practitioners up to five attributes that are most important for requirements engineers. After, we interview a subset of those who responded the survey to gather more information about the mentioned attributes. Lastly, we asked the interviewed participants about the level of importance of the identified strategies to pursue the attributes of a great requirements engineer. The level of importance is related to the extent to which a strategy can support a requirements engineer to improve her/his skills to get closer to being a great requirements engineer.

We surveyed 18 Brazilian software practitioners and 11 of them accepted our invitation for an interview. Based on their answers, this paper's contributions include (i) a list of 22 attributes of great requirements engineers, (ii) a map which relates those attributes, and (iii) a set of 38 strategies that can be employed to pursue them. Software practitioners can use the map to improve their understanding about the role and responsibilities of requirements engineers. The identified strategies work as tools to improve their skills.

Besides this introduction, this paper is organized in six other sections. Section II presents related work. Then, Section III presents the data collection and analysis procedures. Section IV presents the results of the survey and follow up interview. Section V discusses the main findings. Section VI presents the threats to the study validity. Lastly, Section VII presents our final remarks and the next steps of this work.

## II. RELATED WORK

Much of our knowledge about desirable attributes of software practitioners comes from studying specific roles such as software engineer, project manager, and software manager. Li et al. [4] investigated attributes of great software engineers by interviewing experienced engineers. They reported 53 attributes of great software engineers, which were classified into two categories, internal and external. Examples of identified attributes are improving, knowledgeable about people and the organization, and creative.

In another study, Kalliamvakou et al. [5] performed a two-phase study to investigate the attributes of a great manager of software engineers. By interviewing software practitioners from Microsoft, the authors collected the perception of predefined attributes identified in technical literature or used by Microsoft. Afterwards, a survey was grounded from the interview results to identify the attributes' importance. The authors identified 15 attributes organized into a conceptual framework. Furthermore, the attributes were divided into two dimensions (the levels of interaction and the engineering manager's functions) and categories (team, individual, manager cultivating engineering wisdom, motivating the engineers, and mediating communication). Maintains a positive working environment, grows talent, and enables autonomy were the most important attributes from the point of view of the survey participants.

More recently, Dias et al. [6] investigated the attributes of great open-source software maintainers. By conducting an interview study with open-source software maintainers, the authors identified 22 attributes. After, by surveying contributors of open-source software, the authors collected qualitative information about the importance of these attributes. Communication, quality assurance, and community building are considered the most important attributes. The authors also organized the attributes in a conceptual framework, showing the relationship between them.

Lastly, Gren and Ralph [7] surveyed agile professionals to identify which makes effective leadership in agile software teams. The authors identified ten categories (e.g., step in if needed, build a strong team identity, and adapt to customer culture) divided into three themes (dynamic team leadership, social identity, and organizational culture). These categories and themes were organized in a model of agile leadership.

We recognize the valuable results reported by related work. Indeed, we were inspired by them to define our research method based on survey and follow up interviews (see Section III). However, it is still missing the mapping of attributes of great requirements engineers. This is precisely the topic we approach in this paper.

We compare related work with our results in Section V-B to investigate if our initial guess ("We believe current information about the attributes of great software practitioners might not be easily translated to the context of requirements engineering") holds.

## III. RESEARCH METHOD

We followed a method based on the work of Kalliamvakou et al. [5] and Dias et al. [6]. For answering the research questions, we surveyed software practitioners to identify the attributes of great requirements engineers (RQ1), interviewed a subset of them to gather more details about each attribute (RQ2 and RQ3), and asked the interviewed participants a follow up question to identify the level of importance of the strategies to become a great requirements engineer (RQ4). The following subsections present the data collection and analysis procedures for each step of the research method.

TABLE I
SURVEY QUESTIONS

| ID | Survey question (SQ) description | Type |
|---|---|---|
| SQ1 | What is your name? | Open-ended |
| SQ2 | Please, inform us your e-mail address, so we can contact you in the next step of the study. | Open-ended |
| SQ3 | How would you describe your gender? | Closed |
| SQ4 | How many years of experience do you have with software requirements? | Open-ended |
| SQ5 | How many companies have you worked for? | Open-ended |
| SQ6 | What is the size of your company (software and other areas)? | Closed |
| SQ7 | What project role are you assigned in this company? | Closed |
| SQ8 | How do you rate your experience in this role? | Closed |
| SQ9 | Based on your experience, what attributes of a software requirements engineer would you rate as the attributes that matter most to being a great requirements engineer? Please enter up to five attributes ordered by their level of importance, with the highest importance listed first. | Open-ended |

TABLE II
CATEGORIES ADAPTED FROM DIAS ET AL. [6]

| Name | Description |
|---|---|
| Management | This category refers to attributes that help managing a project, in the sense of understanding the vision of the project, establishing and communicating project goals, proposing timelines, managing the quality of documentation, etc. |
| Personality | This category refers to how practitioners express what they think, feel, and behave in their interaction with other project members, referring to personality aspects. |
| Social | This category groups attributes related to how practitioners deal with other project members. |
| Technical | This category is about technical skills that are used to perform requirements engineering activities. |

TABLE III
INTERVIEW SCRIPT

| Section | Interview question (IQ) description |
|---|---|
| Opening | We present the informed consent term. **IQ1.** In your words, how would you define a great software requirements engineer? **IQ2.** Have you ever worked with a great software requirements engineer? **IQ3.** Why did you consider him/her a great requirements engineer? |
| Attributes for a great requirements engineer | In the survey, you listed five desirable attributes for a great requirements engineer. Now, let us address three of them: <<attributes a, b, and c>>. **IQ4.** How would you define *attribute a*? **IQ5.** Why is *attribute a* important for great software requirements engineers? **IQ6.** What are the strategies to obtain *attribute a*? **IQ7.** How can a requirements engineer use *attribute a*? **IQ8-IQ11.** Same questions for *attribute b*. **IQ12-IQ15**. Same questions for *attribute c*. |

### A. Survey on Attributes of Great Requirements Engineers

*1) Design and Data Collection:* We invited requirements engineers and project managers from our Brazilian industrial partners. We considered these two roles to capture multiple perspectives: those responsible to performing RE activities and those who manage them.

The survey comprises a set of characterization questions and an open-ended question for participants indicate up to five attributes that are most important for performing RE activities. We asked only up to five attributes because individuals usually pay more attention to the top few choices, mitigating additional noise in lower rankings [12]. Focusing on just five attributes would (i) keep participants from running their response, (ii) require them to provide focused responses, discarding less relevant attributes, and (iii) reduce the cognitive load of remembering specific past experiences. Table I presents the survey questions and their types (closed or open-ended).

*2) Data Analysis:* We performed different data analysis procedures, as the survey is composed of closed and open-ended questions. For closed questions, we calculated the number of participants choosing an option. After, we summarize the participants' characterization. For open-ended questions, we analyzed the list of attributes reported by the participants looking for similarities, i.e., two or more attributes have the same meaning. For example, the attributes understand the business and knowledge of the business problem were unified under *understand the business* attribute. First and second authors analyzed individually the list to standardize the attributes. Divergences were resolved by the last author. This process was necessary because we did not provide a pre-determined list of attributes for the participants.

We also perceived that the attributes are related to each other, allowing us to group them into categories. The categories are based on those defined by Dias et al. [6] (see Table II). For example, we used *social* category to group the attributes *good listener*, *investigative ability in talking to stakeholders*, *clarity*, and *knowledge about user behavior*. The grouping process was performed using constant comparison [13] and was performed by the first author and reviewed by the last author.

### B. Follow up Interview

*1) Design and Data Collection:* We defined the interview script (Table III) comprised of two sections. In the **opening** section, the participant provided his/her definition for great software requirements engineer and described characteristics of a great one that he/she have worked with. In **attributes of a great software requirements engineer** section, the participant defines, explains the importance, gives strategies to achieve, and exemplifies the use of three attributes. Although each participant has informed up to five attributes in the survey, we choose the three highest-ranked attributes from her/his list.

We carried out the interviews remotely. Each of them lasted around 30 minutes and were recorded with the participant's permission.

*2) Data Analysis:* We transcribed the interviews and organized the answers by questions. Then, we coded the transcriptions to identify the main idea presented in each answer [13], [14]. For example, the participant *P1* reported the following answer to IQ1: "the issue of communication is important, because it has to be a person who knows to listen and intervene at the right times (soft skills), and from a technical point of view, it is more important to know the domain." In this example, we found the following codes: *communication*, *listening skills*, *intervention at appropriate time*, *interpersonal skills*, and *domain knowledge*. In IQ6, the respondents indicated strategies to obtain a specific attribute. For example: "knowing how to listen, knowing how to write, passing on the information that the client needs to whoever is developing (P7)." From this response, we identified the following strategies: *knowing how to listen, improving writing, and sharing the customer's need with the development team.* The coding process was performed by the first and second authors. Each of them, individually, coded half of the transcripts and then revised each other. Afterwards, they met with the last author to resolve the divergences.

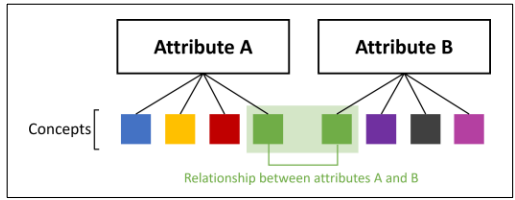

Fig. 1. Analysis Schema

The codes extracted from the definition of attributes reported by participants in IQ4 are **concepts**. We applied conceptual mapping over the concepts to identify relationships between the attributes (see Fig. 1). For example, we identified that the concept *conduct meetings with customers and users* is present in the definition of the attributes *translate user needs into software requirements* and *good knowledge of requirements engineering practices*. Thus, we assumed that these attributes are related to each other. Both of them depends somehow on conduct meetings with customers and users. The first and second authors performed this process, and met with the last author to review the results.

*C. Follow up Question on the Importance of the Identified Strategies to Pursue the Attributes of Great Requirements Engineers*

*1) Design and Data Collection:* We performed this step after obtaining the results from the follow up interviews. To determine the level of importance of each strategy we identified during the interviews (IQ6), we asked a follow up question (see Table IV) to the interviewed participants. For each strategy, we presented its definition and asked the participant to indicate how important it is for a requirements engineer to improve her/his skills to get closer to being a great requirements engineer, considering a 5-point Likert scale (strongly disagree, disagree, neither agree nor disagree, agree, and strongly agree).

*2) Data Analysis:* We used descriptive statistics to analyze the answers by counting the share of participants choosing each option available to rank the strategies.

TABLE IV
FOLLOW UP QUESTIONNAIRE

| ID | Follow up question (FQ) Description | Type |
|---|---|---|
| - | We identified 38 strategies that could be used to enhance the attributes of a great requirements engineer. Below, we present these strategies. For each of them, please indicate how important it is for a requirements engineer to become a great requirements engineer. | - |
| FQ1 | **Participating in training** Improving the skills of the requirements engineer through courses, reading specialized material and getting to know new methodologies. | Closed |
| FQ2-FQ38 | ... | Closed |

IV. RESULTS

In total, we invited 18 Brazilian software practitioners and they completely answered the survey, but only 11 of them were available for interviews and for answering the follow up question. We used answers from the survey (18 answers) to answer RQ1, while the interviews redand answers for the follow up question were used to answer RQ2-3 and RQ4, respectively.

*A. Demographics*

Table V presents the participants' characterization indicating their years of experience, the number of organizations the participant has worked, current company size (small - up to 50 employees; medium-sized - with 51 to 1000 employees; large - more than 1000 employees), and role. All participants are from different organizations.

We can notice that most of the participants are requirements engineers and most have from eleven to 20 years of level of experience in their roles. Most of the participants work in medium-sized companies and most have worked in at least six different companies in their professional life.

Although our sample consists of software practitioners from the Brazilian software industry, it includes a diverse group of professionals with varying levels of experience in project management or requirements engineering roles across companies of different sizes. Details of this diversity can be observed in Table V.

*B. RQ1: What are the Attributes of a Great Requirements Engineer?*

We identified 22 attributes of a great requirements engineer. Table VI presents the five most cited attributes, along with the total number of participants that cite an attribute (#CA) and the percentage of the total of mentions (%A). The complete list of attributes is available in [15]. *Investigative ability in talking to stakeholders* was the most mentioned attribute, followed by *judicious*, *understand the business*, *good ability to identify missing requirements*, and *good knowledge of requirements engineering practices*. These attributes represent 55% of all mentioned attributes frequency.

From the interviews, we identified the definition of the attributes. *Investigative ability in talking to stakeholders* attribute means that a software requirements engineer must be

TABLE V
Participant's demographics

| Participant ID | Years of Experience | #Companies worked | Company Size | Role | Follow up Interviews (Y/N) |
|---|---|---|---|---|---|
| P1 | 15 | 4 | Large | Project manager | Yes |
| P2 | 2 | 3 | Small | Requirements engineer | Yes |
| P3 | 10 | 6 | Large | Requirements engineer | Yes |
| P4 | 25 | 5 | Large | Project manager | Yes |
| P5 | 3 | 4 | Small | Requirements engineer | Yes |
| P6 | 2 | 3 | Medium-sized | Requirements engineer | Yes |
| P7 | 16 | 7 | Large | Requirements engineer | Yes |
| P8 | 5 | 10 | Large | Project manager | Yes |
| P9 | 5 | 8 | Medium-sized | Requirements engineer | Yes |
| P10 | 30 | 11 | Large | Project manager | Yes |
| P11 | 10 | 11 | Medium-sized | Project manager | Yes |
| P12 | 15 | 5 | Large | Project manager | No |
| P13 | 1.5 | 1 | Medium-sized | Requirements engineer | No |
| P14 | 15 | 3 | Large | Project manager | No |
| P15 | 12 | 5 | Large | Requirements engineer | No |
| P16 | 1 | 1 | Medium-sized | Requirements engineer | No |
| P17 | 20 | 10 | Large | Project manager | No |
| P18 | 1 | 4 | Medium-sized | Requirements engineer | No |

TABLE VI
The five most cited attributes

| Attribute | #CA | %A |
|---|---|---|
| Investigative ability in talking to stakeholders | 15 | 18% |
| Judicious | 12 | 14% |
| Understand the business | 7 | 8% |
| Good ability to identify missing requirements | 7 | 8% |
| Good knowledge of requirements engineering practices | 5 | 6% |

Caption:
#CA: Number of mentions of an attribute.
%A: Percentage of #CA in relation to the total all mentioned attributes (83).

prepared not to limit themselves to what the interested party is saying, but to try to provoke them for answers about the system requirements, as described by the participant P3 "so you have to prepare, study, create a script, so I think that this investigative capacity comes from there..." *Judicious* attribute refers to the ability to organize the software documentation and pay attention to details, for instance, P6: "As a very detail-oriented person, he not only investigate but detail the information obtained so that it becomes clear in documentation and easy to communicate."

*Understand the business* attribute is related to understand the environment where the system will be used, as described by P1: "It would be the understanding of the business context, where that system will be inserted. Nothing from a technical point of view, from the business itself, from the environment..." *Good ability to identify missing requirements* attribute is associated with the ability to propose solutions considering the user context and validate them with customers, as illustrated by P5: "the analyst needs to be able to understand when something is not clear, when things are not going well. And after, having the ability to elicit what is missing and the ability to validate that with the customer." Lastly, *good knowledge of requirements engineering practices* attribute refers to have knowledge on the practices used to elicit, analyze, and specify software requirements. For instance, participant P1 reported "understanding the technique behind these elements that we use... to get in touch with users..."

By constantly comparing the codes [13], we grouped them into the following categories:

- **Personality**: It has seven attributes to express how requirements engineers think, feel, and behave in their interaction with stakeholders. These attributes are: *agility*, *analytical reasoning*, *antifragility*, *associative reasoning*, *commitment*, *curiosity*, and *judicious*.
- **Management**: It groups the following attributes to support managing a project or the requirements: *good ability to identify missing requirements*, *good systematic view*, *understand the business*, *update requirements (when necessary)*, and *set priorities with customer*.
- **Technical**: It is composed of seven attributes associated with technical skills that are used in software requirements activities. These attributes are: *ability to validate implemented requirements*, *good writing (knowing how to write clearly and without ambiguities)*, *experience*, *good knowledge of general software development practices*, *good knowledge of requirements engineering practices*, *prior knowledge, even if basic, in some programming language*, and *translate user needs into software requirements*.
- **Social**: It encompasses the following attributes related to how requirements engineers deal with stakeholders: *clarity*, *good listener*, *investigative ability in talking to stakeholders*, and *knowledge about user behavior*.

Table VII presents the categories of attributes, reporting the category's name, the number of unique attributes mentioned (#A) and the total number (i.e., count) of participants that cite attributes in each category (#CA). Lastly, the column %A corresponds to the percentage of #CA in relation to all mentioned attributes. The categories personality and social are the most mentioned, highlighting the central skills of a great requirements engineer.

*1) Difference in perception of the attributes between requirements engineers and project managers:* As our popula-

TABLE VII
CATEGORIES OF ATTRIBUTES

| Category | #A | #CA | %A |
|---|---|---|---|
| Personality | 7 | 25 | 30% |
| Social | 4 | 24 | 29% |
| Management | 5 | 18 | 22% |
| Technical | 6 | 16 | 19% |

Caption:
#A: Number of unique mentioned attributes.
#CA: Number of mentions of an attribute.
%A: Percentage of #CA in relation to the total all mentioned attributes (83).

tion is composed of requirements engineers and project managers, we investigated whether they have a difference in their perception of the attributes of a great software requirements engineer. Figure 2 shows a Venn diagram representing the result of the comparison. We identified six attributes only reported by project managers. These attributes are related to soft skills that increase the requirements engineer's ability to perform their activities. We also recognized four attributes reported only by requirements engineers. These attributes are associated with hard skills needed to perform software requirements activities. Finally, twelve attributes were reported by both.

*C. RQ2: How these attributes are related to each other?*

In total, we identified 13 relationships between attributes. Table VIII shows these relationships along with their attributes, concepts, and categories. A concept is the definition part shared by two or more attributes. For example, the concept *be investigative to define and associate the problems* is part of the definition of the following attributes *analytical reasoning*, *associative reasoning*, and *judicious*, as we can see in "P8: It's the ability you have to see a problem. And being able to find requirements to define that problem," "P8: It's you being able to associate a problem you're seeing today, and problems you've seen in the past," and "P6: As a very detail-oriented person, he not only investigate but also detail the information obtained so that it becomes clear," respectively. In Table VIII, we only show the attributes that have at least one relationship with other attribute.

We organized these concepts and the relationships in a map (Fig. 6) which is further discussed in Section V-A.

*D. RQ3: What strategies can be used to obtain these attributes?*

From the interviews, we identified 38 strategies to obtain the attributes of a great requirements engineer. Table IX presents the five most cited strategies, along with the total number (i.e., count) of participants that cite a strategy (#CS) and the percentage of the total of mentions (%S). The complete list of strategies is available in [15]. The most mentioned strategy was blue*Participating in training*, followed by *acquiring domain knowledge*, *talking to all interested parties*, *searching for solutions that fit the user's reality*, and *acquiring knowledge about tests*.

*Participating in training* means improving the skills of the requirements engineer through courses, reading specialized material and getting to know new methodologies. For instance, "P1: Theoretical knowledge, which teaches how to apply the techniques, what is relevant, what kind of attitude the analyst has to have, how he has to prepare, the points of attention he has to have." *Acquiring domain knowledge* refers to knowledge about a field for which the system is under developed, as described in "P1: studying that area in which the system is inserted" and "P5: knowing your (system) domain."

*Talking to all stakeholders* strategy is related to keep in touch with customers, users, and software teams to understand and explain the requirements, such as "P7: Talking, demonstrating their ideas, talking both with the customer and the software developers, listening, exchanging information." *Searching for solutions that fit the user's reality* strategy is associated with defining a solution considering users' needs and restrictions, as illustrated in "P9: it is necessary to seek an innovative, low-cost solution that is developed in the shortest possible time." Lastly, *Acquiring knowledge about tests* means the ability to define test cases to verify the system in relation to its requirements, for example, the participant P11 reported "...create test objectives, create test sections..., know test techniques."

We also investigated the relationships between the strategies and attributes, organizing them in a Sankey diagram [16]. This diagram is composed of bars, representing the sources and destination of information, and links, showing the magnitude of the flow between those bars. Figure 3 shows the diagram for the five most mentioned attributes and the three most mentioned strategies to obtain them. A complete version is available in [15]. The flow starts with attributes which is related to strategies. The numeric values appear with each element (attribute and strategy) show the total number of times that a relationship occurs. The thickness of each link varies according to the value of the relationship. For example, the *participating in training* strategy can be used to obtain the attributes *understand the business*, *good knowledge of requirements engineering practices*, and *judicious*.

*E. RQ4: How practitioners perceive the importance of the identified strategies to pursue the attributes of great requirements engineers?*

Figure 4 presents the results from the follow up question answered by all interviewees. In general, the strategies were well evaluated. *Participating in training*, *acquiring domain knowledge*, *seeking know hear*, *seeking a better communication*, *improving writing*, *being patient*, and *acquiring knowledge on requirements analyses* were considered the main strategies to improve the ability of requirements engineers. On the other side, *searching for similar systems/solutions*, *defining a folder structure to quickly identify documents on a specific subject*, *using HCI techniques*, *using WIKI for sharing requirements documentation*, and *reading source code* are only mentioned by some of the participants.

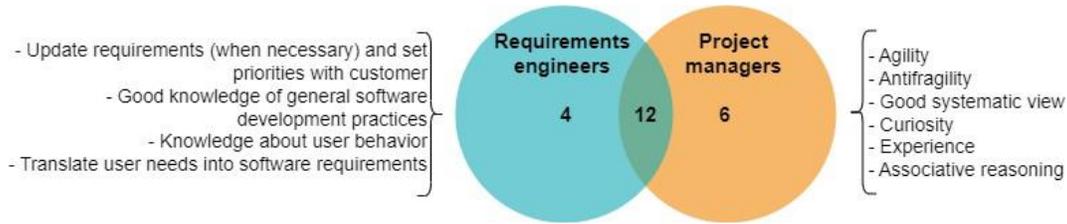

Fig. 2. Venn diagram comparing attributes from requirements engineers' and project managers' perspective

TABLE VIII
RELATIONSHIPS BETWEEN THE ATTRIBUTES

| Category | Attribute relationships | Concept |
|---|---|---|
| Management | Good ability to identify missing requirements ↔ Understand the business | Ability to understand the information flow |
| | Good ability to identify missing requirements ↔ Understand the business | Ability to understand user regulation |
| | Good ability to identify missing requirements ↔ Understand the business | Ability to understand business concept |
| Personality | Associative reasoning ↔ Agility | Turn around problems quickly |
| | Agility ↔ Judicious | Keep information up to date by understanding the business |
| | Analytical reasoning ↔ Associative reasoning ↔ Judicious | Be investigative to define and associate the problems |
| Social | Clarity ↔ Investigative ability in talking to stakeholders | Refine the collected information |
| | Clarity ↔ Knowledge about user behavior | Refine the collected information |
| | Good listener ↔ Investigative ability in talking to stakeholders | Do not interfere with the customer's speech |
| | Good listener ↔ Investigative ability in talking to stakeholders | Do not think you know everything |
| | Good listener ↔ Investigative ability in talking to stakeholders | Listen to extract the most information |
| Technical | Translate user needs into software requirements ↔ Good knowledge of requirements engineering practices | Conduct meetings with customers and users |
| | Translate user needs into software requirements ↔ Good knowledge of requirements engineering practices | Apply requirements elicitation techniques to understand the users' needs |
| | Good knowledge of general software development practices ↔ Experience | Know the areas of RE |

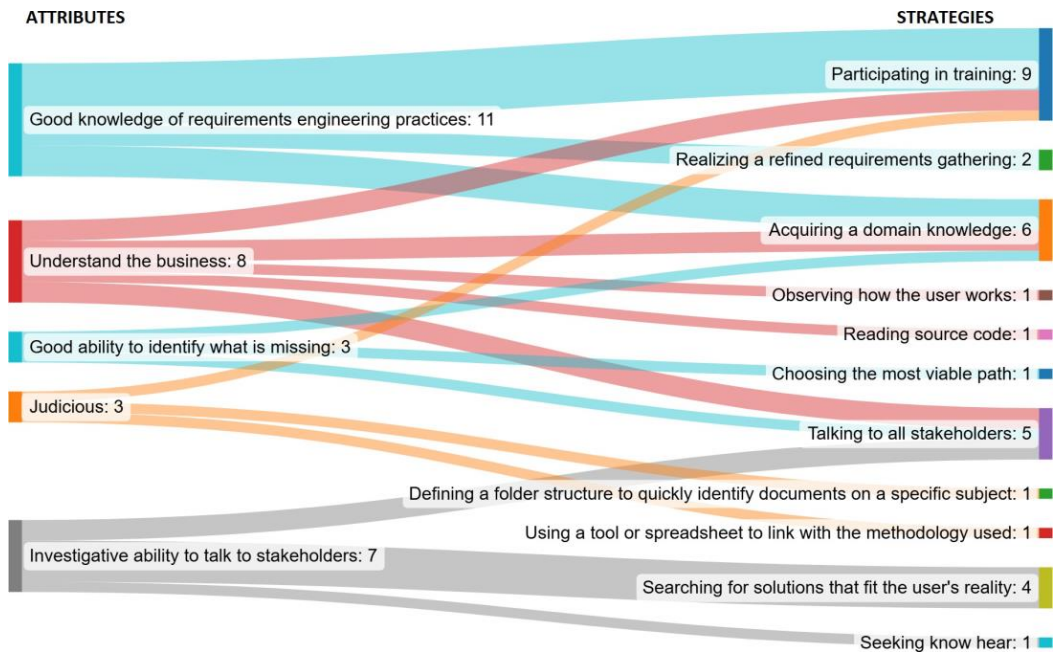

Fig. 3. Sankey diagram for the most mentioned attributes and the top three strategies to obtain them

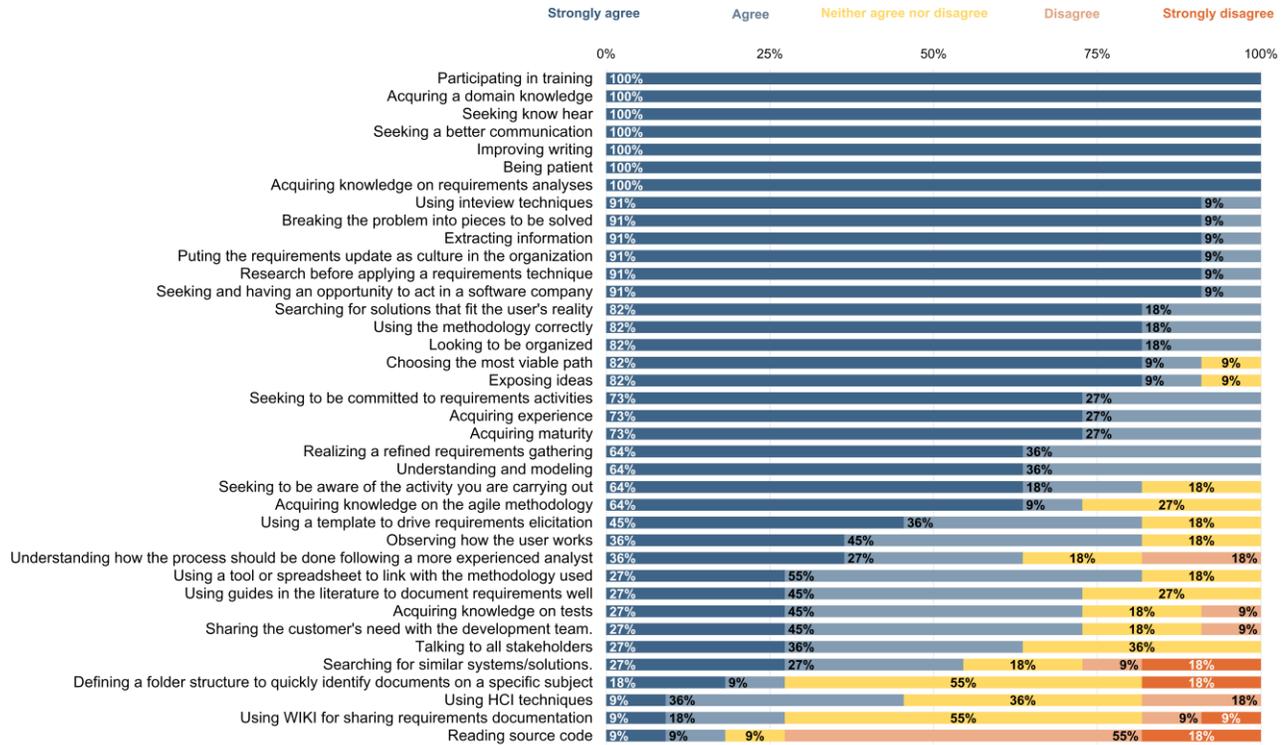

Fig. 4. Interviewees' point of view on the importance of the identified strategies

TABLE IX
THE FIVE MOST CITED STRATEGIES

| Strategy | #CS | %S |
|---|---|---|
| Participating in training | 17 | 19% |
| Acquiring domain knowledge | 9 | 10% |
| Talking to all stakeholders | 7 | 8% |
| Searching for solutions that fit the user's reality | 4 | 4% |
| Acquiring knowledge on tests | 4 | 4% |

**Caption**:
#CS: Number of mentions of a strategy.
%S: Percentage of #CS in relation to the total all mentioned strategies (91).

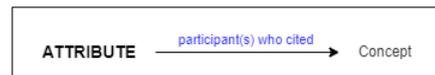

Fig. 5. Elements of the map

## V. DISCUSSION

This section presents the map that organizes the attributes and their concepts and relationships. Also, it compares our findings with related work and discusses the implications of our results for researchers, managers, novice requirements engineers, and educators.

### A. Map of Great Requirement Engineers

We organized the set of concepts related to each attribute and their relationships in a map. Inspired by the framework structure proposed by Dias et al. [6], attributes are represented in bold upper-case text, whereas the concepts are represented as smaller, lowercase text. An attribute has a set of concepts that describe it. An arrow relates an attribute and its concepts, and has a blue text to indicate the participant who mentioned the concepts. Fig. 5 exemplifies how our map is structured.

Fig. 6 presents the four maps. We define a map for each attribute category. We can notice that all attributes from management (A) category is related to at least one other attribute. From personality (B) and technical (D) categories, we found relationships for almost all attributes, except *commitment* and *good knowledge of general software development practices* attributes. In the social (C) category, the attributes *clarity*, *investigate ability in talking to stakeholders*, and *know about user behavior* are related to each other. Lastly, the maps do not present *antifragility*, *good systematic view*, *prior knowledge, even if basic, in some programming language*, and *curiosity* attributes, because they were not covered in the interviews.

We also found core concepts (highlighted in green), i.e., a concept that relates two or more attributes. For example, *well-defined and updated documentation* (from Management category) is a concept that relates two attributes *update requirements (when necessary) and set priorities with customer*, and *understand the business*.

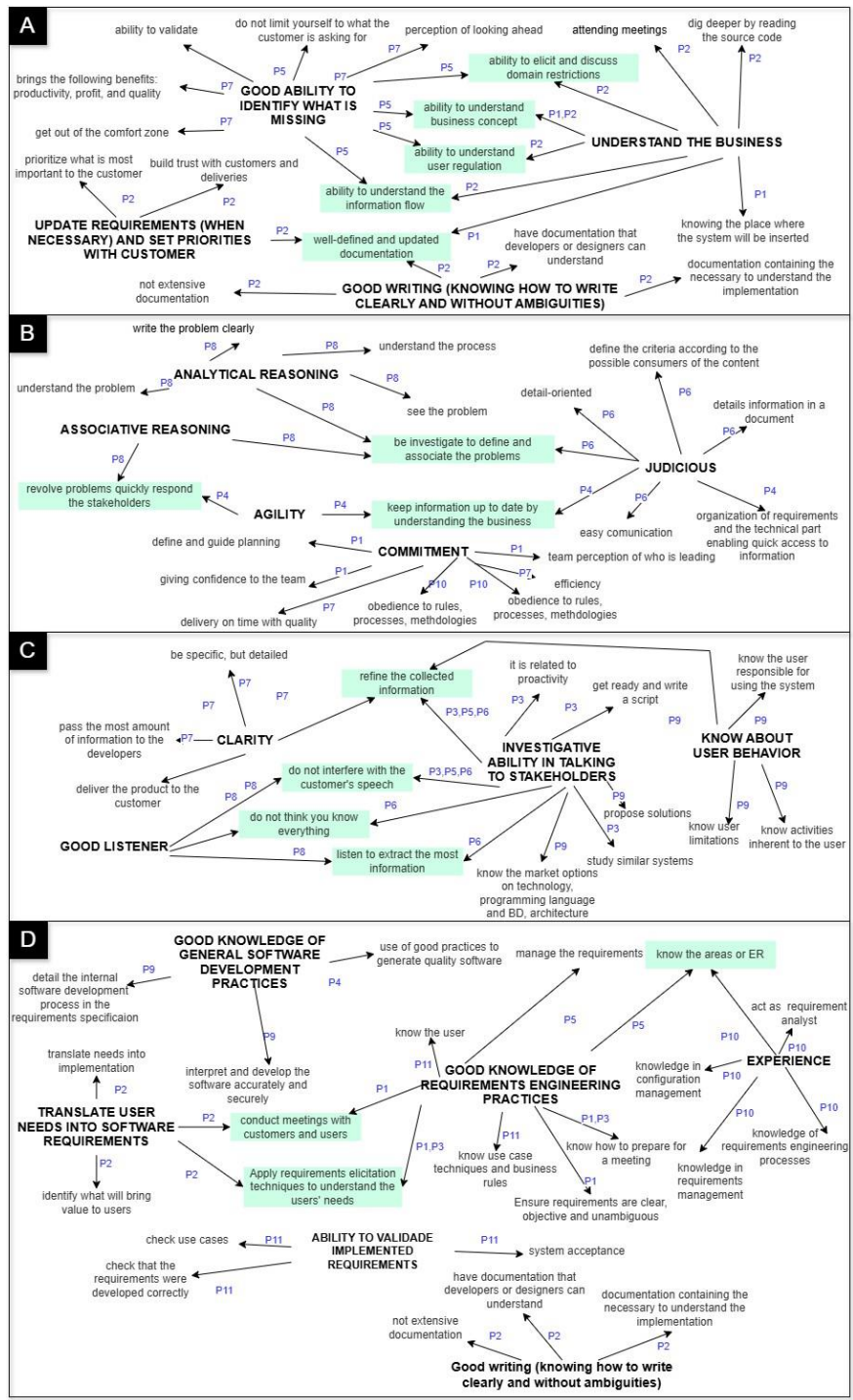

Fig. 6. Map for (A) management, (B) personality, (C) social, and (D) technical categories.

Software professionals can use the map to begin or improve their initiatives to address these attributes. As a reference, software professionals can understand how attributes are defined and seek strategies or improvements to achieve or improve these attributes in their projects. As a conceptual guide, the map can be employed to inform actions in response to perceived attributes, and as a comprehensive guide to evaluate strategies or improvements that can be achieved so that these attributes are increasingly present in our team. It facilitates the identification and recognition of attributes (represented in bold capital text) and concepts related to each attribute (represented in smaller, lowercase text).

### B. Comparison to Related Work

Table X presents a comparison between our results, text books (columns in gray), and related work (columns in blue). Considering the attributes described in the text books [8], [9], we can see that the common attributes are more related to human-aspects, such as *analytic thinking*, *responsibility*, *communication*, and *persuasiveness*. The other attributes found in our study reveal other abilities that requirements engineers have to pursue to develop their activities.

Concerning related work (columns in blue), the common attributes between a great requirements engineer and a great software engineer [4] are related to soft skills (*agility*, *commitment*, *curiosity*, *good systematic view*, and *judicious*) and knowledge on processes, practices, and domain (*good knowledge of general software development practices*, *good knowledge of requirements engineering practices*, and *understand the business*). The other attributes found only for requirements engineers are associated with specific requirements activities such as *good listener*, *good writing (knowing how to write clearly and without ambiguities)*, and *translate user needs into software requirements*.

We also found a few common attributes with great open-source software maintainers, managers of software engineers, and leadership in agile software teams.

In summary, results confirmed our initial guess that attributes of great software practitioners might not be easily translated to the context of requirements engineering. Our results complement the set of information already reported in technical literature, providing specific attributes to requirements engineers.

### C. Implications

The attributes we have identified and described may have wide-ranging implications for researchers, managers, novice requirements engineers, and educators. In the rest of this section, we discuss implications and opportunities to build upon our results.

*1) For Researchers:* Researchers can consider the identified attributes in the development of new methods, strategies, and tools that help requirements engineers in carrying out their activities and, by making use of them, achieve the desirable attributes of a great requirements engineer.

Researchers might also investigate interventions that help achieve the attributes quickly and effectively.

*2) For Managers:* The map can be used to support managers to more effectively communicate attributes of a great requirements engineer. As a conceptual guide, the proposed map along with the Sankey diagram facilitate a more effective identification, acknowledgement and relationship of attributes of great requirements engineers. Software practitioners can also use them to become better requirements engineers. For instance, managers can identify weak points in their teams and define strategies to reduce or even eliminate those weaknesses. If a manager identifies that the attribute *good knowledge of requirements engineering practices* is not ideal in her/his team, analyzing the relationships of this attribute, he/she can realize that by pursuing the concepts managing the requirements and knowing the requirements engineering areas he/she will be able to improve her/his team's skills.

*3) For Novice Requirements Engineers:* It is a common issue to see new requirements engineers unsure of how to become great requirements engineers. The results reported in this work list a series of attributes that they might pursue. Moreover, our results provide a set of strategies that can be employed to this end. Improvements might come, for instance, from participating in training, acquiring domain knowledge, or talking to all stakeholders.

Our findings may also help new requirements engineers better present themselves in job searching. Novice engineers can consider demonstrating to employers that they have or can develop these attributes.

*4) For Educators:* Our results raise questions about curriculum choices, teaching methods and learning objectives in software engineering education. Educators may consider adding courses on topics not found in their current curricula. For example, a specific course on requirements engineering could be necessary to make possible the teaching of requirements engineering practices.

Software engineering educators might also use our results to review their teaching method. Many attributes of great requirements engineers are soft skills. Teaching methods based on experiential learning practices could be useful to improve students' soft skills [17]. In experiential learning, students actively engage in complex tasks that reflect the problems they likely encounter in the workplace. To bridge the academic and employment skills and knowledge they are developing through these tasks, students participate in reflective activities that help them articulate the relevance and implications of the experience for lifelong learning [17].

Lastly, educators might also explicitly discussing what students will not learn in school, allowing them to be aware of potential knowledge gaps, preparing them to search for opportunities outside of the academic setting (e.g. internships). For example, attributes like *investigative ability in talking to stakeholders* may not be reasonable to teach in an academic setting.

## VI. THREATS TO VALIDITY

As with any empirical study, there may be threats and/or limits to our methods and findings [18].

TABLE X
COMPARISON WITH RELATED WORK

| Great requirements engineer (our study) | Attributes for a | | | | | |
|---|---|---|---|---|---|---|
| | Requirements engineer [8] | Requirements engineer [9] | Great software engineer [4] | Great manager of software engineers [5] | Great open-source software maintainer [6] | Great leadership in agile software teams [7] |
| Ability to validate implemented requirements | - | - | - | - | - | - |
| Agility | - | - | Productive | - | - | - |
| Analytical reasoning | Analytic thinking | - | - | - | - | - |
| Antifragility (it is a concept that emphasizes the ability to evolve in the face of adverse situations) | - | - | - | - | Diligence | - |
| Associative reasoning | - | - | - | - | - | - |
| Clarity | - | - | - | - | - | - |
| Commitment | - | Responsibility | Hardworking | - | Responsibility | Step in, if needed |
| Curiosity | - | - | Curious | - | - | - |
| Experience | - | - | - | - | - | - |
| Good ability to identify what is missing | - | - | - | - | - | - |
| Good knowledge of general software development practices | - | - | Knowledgeable about engineering processes | Is technical | - | - |
| Good knowledge of requirements engineering practices | - | - | Knowledgeable about their technical domain | - | Technical excellence | - |
| Good listener | Communication skills | Communication | - | - | - | - |
| Good systematic view | - | - | Systematic | - | - | - |
| Good writing (knowing how to write clearly and without ambiguities) | Communication skills | Communication | - | - | - | - |
| Investigative ability in talking to stakeholders | Persuasiveness | - | - | - | Communication | - |
| Judicious | - | - | Attentive to details | - | Discipline | - |
| Knowledge about user behavior | - | - | - | - | - | - |
| Prior knowledge, even if basic, in some programming language | - | - | - | - | - | - |
| Translate user needs into software requirements | - | - | - | - | - | - |
| Understand the business | - | - | Knowledgeable about customers and business | - | Domain Experience | Understanding the company ecosystem |
| Update requirements (when necessary) and set priorities with customer | - | - | - | - | - | - |

**External validity**: Our analysis comes wholly from 18 software practitioners from the software development industry from one country (Brazil). This makes it unlikely that our results are representative of the views of software managers and requirements engineers in general. To minimize this threat, we considered a variety of participants in terms of years of experience, company size, and number of companies worked. Besides, Brazil is a continental-sized country, the largest in South America, presenting enormous landscape, economic and cultural diversity. An argument can be made, however, that ecological validity [19] of the work (i.e., the extent to which these findings approximate other real-world scenarios) does hold. Another threat that may affect external validity may be related to how many of these individuals have been exposed to truly great software engineers, instead of a number of more-or-less competent ones. To mitigate this threat, participants should justify why they consider the software engineers to be a great software engineer based on the attributes they define. In addition, we may have the fact that our domain is returning to the area of information systems. And to mitigate this threat we consider different domains (such as, commercial and financial).

**Internal validity**: Threats affecting the internal validity can arise from the questionnaire. As it was applied remotely, the participants could misinterpret the questions. To mitigate this threat, the questionnaire was submitted to two internal and one external validations, and a pilot study. Another threat is related to the term great requirements engineer used in the questions. As the participants could misinterpret this term, they could give invalid answers. To reduce this threat, we analyzed all answers given to SQ9 and concluded that no invalid answer was reported.

**Conclusion validity**: A threat arises from the qualitative analyses we carried out because they are a subjective task. To mitigate this threat, the analyses were carried out separately by two researchers and the consensus was performed by an experienced researcher. We also used the constant comparison technique when coding the interview transcripts, comparing our findings with previous ones as they emerged from the data analysis. Another threat affecting the conclusion validity is related to the participants' project role. Although we recognize the importance of requirements engineers in the context of this work, we did not survey only practitioners performing requirements tasks. Having a set of participants encompassing managerial tasks is also necessary because we can further understand the attributes of a great requirements engineer, as demonstrated in Subsection IV-B1.

## VII. CONCLUDING REMARKS

In this paper, we investigate the attributes of a great requirements engineer. We organize these attributes into a map that can support software professionals in seeking strategies to become better requirements engineers. Also, the map can guide new research efforts in a problem-driven way.

Our results also open many opportunities of future work. Examples of research gaps implications of the results for supporting requirements engineers, managers and educators are discussed in Section V-C. Particularly, as next steps of this research, we intend to (i) conduct a rating survey with requirements engineers and project managers aiming at understanding how they prioritize the identified attributes, (ii) investigate undesirable attributes of a requirements engineer, and (iii) investigate teaching strategies that are suited to educate new requirements engineers considering both technical and soft skills.

## ACKNOWLEDGMENTS

The authors would like to thank all software practitioners who participated in the experimental studies. This study was financed in part by the Coordenação de Aperfeiçoamento de Pessoal de Nível Superior Brasil (CAPES) - Finance Code 001